 \documentclass[preprint]{aastex}

\slugcomment{PASP, in press.   \qquad 17 Oct. 2001}

\shorttitle{JHKL$'$M$'$ Filter Set}
\shortauthors{Tokunaga et al.}

\begin{document}


\title{The Mauna Kea Observatories Near-Infrared Filter Set. II. \\
Specifications for a New $JHKL'M'$ Filter Set for \\
Infrared Astronomy}

\author{A. T. Tokunaga}
\affil{Institute for Astronomy, University of Hawaii,
    Honolulu, HI 96822}
\email{tokunaga@ifa.hawaii.edu}

\author{D.A. Simons}
\affil{Gemini Observatory, Northern Operations Center, 670 N. A`ohoku 
Place, \\Hilo, HI  96720}
\email{dsimons@gemini.edu}

\and

\author{W. D. Vacca}
\affil{Max-Planck-Institut f\"{u}r extraterrestrische Physik\\
         Postfach 1312, D-85741 Garching, Germany}
\email{vacca@mpe.mpg.de}

\begin{abstract}
 We present a description of a new 1--5 $\mu$m filter set similar to
 the long-used $JHKLM$ filter set derived from that of Johnson.  The
 new Mauna Kea Observatories Near-Infrared (MKO-NIR) filter set is
 designed to reduce background noise, improve photometric
 transformations from observatory to observatory, provide greater
 accuracy in extrapolating to zero air mass, and reduce the color
 dependence in the extinction coefficient in photometric reductions. 
 We have also taken into account the requirements of adaptive optics
 in setting the flatness specification of the filters.  A complete
 technical description is presented to facilitate the production of
 similar filters in the future.

\end{abstract}

\keywords{ infrared: general --- instrumentation: photometers}

\section{Introduction}


The rationale for a new set of infrared filters was presented by
Simons \& Tokunaga (2001; hereafter Paper I).  The goals of the design
of this new filter set are similar to those of Young, Milone, \& Stagg
(1994), namely to construct a filter set that minimized color terms in
the transformations between photometric systems and that reduced the
uncertainty in determining the absolute calibration of photometric
systems.  However, in contrast to the filters proposed by Young et
al., the filters in this new set were also designed to maximize
throughput, in addition to minimizing the effects of atmospheric
absorption.  The reduced dependence on atmospheric absorption also
permits photometry to be less sensitive to water vapor variations and
to the altitude of the observatory.  Most importantly, the new filters
permit extrapolation to zero air mass with small errors.

We required large-size, high-quality filters for several facility
instruments for the Gemini and Subaru Telescopes.
Since the cost of each filter is dominated by the
technical difficulties of the coating process it is desirable to place
as many substrates as possible into the coating chamber in order to
reduce the cost per filter.  Hence there is a strong economic driver
to produce custom filters in a consortium production run.  This has
the added benefit that with more observatories involved there would be
greater standardization among the observatories.


Although this filter set was designed for the Gemini and Subaru
Telescopes, all of the optical/infrared observatories at Mauna Kea are
presently using these filters (NASA Infrared Telescope Facility,
United Kingdom Infrared Telescope, Canada-France-Hawaii Telescope,
Keck, Gemini, Subaru).  In addition, the filter set was discussed
informally by the IAU Working Groups on IR Photometry and on Standard
Stars at the 2000 General Assembly, and endorsed as the preferred
``standard'' near-infrared photometric system, to be known as the
Mauna Kea Observatories Near-Infrared (MKO-NIR) photometric system.

In this paper we describe a set of filters that were fabricated
according to the definitions presented in Paper I. We include a list
of specifications so that similar filters can be produced in the
future.  It is our hope that if enough observatories adopt these
filters, we will have greater uniformity among photometric systems, as
well as reduced systematic errors when comparing observations from
different observatories.

\section{Filter Specifications}

We present here the list of technical specifications that were
required to be satisfied by the filter manufacturer.  The center,
cut-on, and cut-off wavelengths are given in Table 1, and they follow
the filter definitions given in Paper I. The specifications for
substrate flatness, parallelism of the filters, and use of a single
substrate are required for use with adaptive optics systems.  

\begin{enumerate}
    
\item {\em Out-of-band transmission: $<$10$^{-4}$ out to 5.6 $\mu$m.} 

This specification is required for use with InSb detectors.  For
HgCdTe 2.5 $\mu$m cut-off detectors, the out-of-band blocking can be
specified for wavelengths less than 3.0 $\mu$m instead of 5.6 $\mu$m. 
The desired blocking is better than 10$^{-4}$, but practical
considerations such as cost and manufacturing difficulty make it
impossible to go lower.  In cases where blocking to this level or
better is not possible, a separate blocking filter should be used. 
PK-50 is a suitable blocker for wavelengths less than 2.0 $\mu$m.  For
longer wavelengths, an interference blocking filter may be required
for InSb detectors.

\item {\em Operating temperature: 65 K.  Cold filter scans of witness 
   samples to be provided, together with prediction of wavelength shift 
   with temperature.}
   
The filters are expected to be used at 50--77 K, but the specified
filter temperature was set at 65 K primarily because the Gemini and
Subaru instruments were designed for use with InSb detectors and with
cryogenic motors inside  the cryostat.  This requires cooling to
65 K to avoid excess thermal emission from the motors 
within the cryostat.

\item {\em Average transmission: $>$80\%  (goal $>$90\%).}  

\item {\em Transmission between the 80\% points to have a ripple 
of less than $\pm$5\%.}

\item {\em Cut-on and cut-off:  $\pm$0.5\% wavelength error.}  

\item {\em Roll-off: \%slope $\leq$2.5\%.}                   

The \%slope $=$
$[\lambda(80\%)-\lambda(5\%) ] / \lambda(5\%) \times 100$, where
$\lambda(80\%)$ is the wavelength at 80\% transmission and
$\lambda(5\%)$ the wavelength at 5\% transmission.

Ideally the filters should have a square ``boxcar'' shape for maximum
throughput.  The specifications on the average transmission, ripple,
cut-on and cut-off uncertainty, and roll-off uncertainty are a
compromise between the desired sharp edge, practical manufacturing
specifications, and cost. 

\item {\em Substrate surfaces parallel to $\leq$5$''$.}           

\item {\em Substrate flatness: $<$$0.0138 \lambda /(n-1)$, where  
   $n$ is the  index of refraction of the substrate (compatible with 
   AO systems).}

For example, for $n=1.5$, $\lambda=2200$ nm, the substrate flatness 
should be $<$61 nm.  For $n=3.4$, $\lambda=2200$ nm, the substrate 
flatness should be $<$13 nm.

This is a flatness specification, not a roughness specification.  The
Strehl ratio (SR) can be approximated as SR $\sim 1 - [ ( 2 \pi /
\lambda ) w ]^{2}$, where $w$ is the rms wavefront error (see Schroeder
1987).  This expression is valid for a single reflective surface.  For
two surfaces in transmission, we have SR $\sim 1 - [ (n - 1) (2 \pi /
\lambda) w \sqrt{2} ]^{2}$, where $n$ is the index of refraction of
the substrate.  For SR = 0.985, $ w = 0.01378 [\lambda / (n-1)]$. 
Alternatively, one could specify $\lambda/10$ peak-to-valley at 0.63
$\mu$m, which is approximately $\lambda/40$ rms, or $\sim$15 nm rms. 
This would be suitable for glass substrates ($n=1.5$) at 1.0--2.2
$\mu$m as well as silicon substrates ($n = 3.4$) at 2.2--4.8 $\mu$m. 
It should be noted that after coating, the substrate will deform under
the stress of the coatings, taking on a concave or convex shape. 
However, this does not have any effect on the wavefront error in
transmission since the surfaces are parallel.

\item {\em Single-substrate to be used (cemented filters not      
   acceptable).}

Cemented filters are not considered acceptable because of potential
large surface deformations when cooled.  In cases where a high SR is
not needed, cemented filters may be acceptable.
   
Items 7--9 are the main requirements for filters to be used in adaptive
optics.  The specified flatness conforms to an SR of 0.985.

\item {\em Filter to be designed for a tilt of 5$^{\circ}$ (to   
   suppress ghost images).}
   
Items 9--10 assume that these filters will be placed at the pupil image
within a camera and the filters are tilted to the optical axis by
5$^{\circ}$ to avoid ghost images in the focal plane.  Alternatively,
these filters could be designed to be used with no tilt, but with a
wedged substrate.

\item {\em Scratch/Dig:  40/20.}                               

This item specifies the maximum size of scratches and digs (pits) that
are permitted on the surface of the substrate.  The specification
40/20, which is acceptable in
most ground-based astronomy applications,  requires that
scratches be no wider than 4 $\mu$m and that the pits be no wider
than 200 $\mu$m in diameter.  A technical description of this
specification is given by Bennett \& Mattsson (1999).

\item {\em Diameter:  60 mm, or cut to requested size.}      

The coating is typically not good near the edge and some of the filter
is blocked by the filter mount.  Therefore the filter diameter is
typically specified to be about 10\% larger than the actual size
required by the optics.

\item {\em Maximum thickness: 5 mm, including blocker for InSb  
   detectors.}
   
Items 12--13 depends on the application.  The
thickness/diameter ratio must be of a practical value for polishing
and mechanical strength and is usually specified to be about 0.1.

\item {\em No radioactive materials such as thorium to be used  
   (to avoid spurious detector noise spikes).}

\end{enumerate}

\section{Filter Fabrication and Filter Profiles}

A number of vendors were contacted to solicit bids on the filter 
production for the consortium of buyers.  The vendor selected was OCLI 
of Santa Rosa, California.  This selection was determined both by 
price of production, ability to meet the specifications, maturity of 
the production facilities, and willingness to accept individual orders 
from the consortium.  The filters were cut to size by the vendor for 
each order.

A production run was ordered to accommodate the filter needs of
the following observatories and institutions:
Anglo-Australian Observatory, 
California Institute of Technology, 
Canada-France-Hawaii Telescope, Carnegie Institution, 
Center for Astrophysics, Cornell University, 
European Southern Observatory, 
Gemini Telescopes, Kiso Observatory,
Korean Astronomy Observatory, Kyoto University,
MPE-Garching, MPI-Heidelberg, 
NASA Goddard Space Flight Center, 
NASA Infrared Telescope Facility, 
National Astronomical Observatory of Japan, 
National Optical Astronomy Observatories, 
Nordic Optical Telescope, 
Ohio State University, Osservatorio Astrofisico di Arcetri, 
Subaru Telescope, United Kingdom Infrared Telescope,  
University of California Berkeley, 
University of California Los Angeles, Univ. of Cambridge,
University of Grenoble, Univ. of Hawaii, 
University of Wyoming, 
and the William Herschel Telescope.

Figures 1(a)--(g) show the transmission profiles of the filters 
that were produced.  The profiles are shown at 65 K, as obtained 
by extrapolating the filter shift with temperature from room temperature 
to 65 K.  Figure 2 shows the $J, H, K_s, L'$, and $M'$ filter 
profiles superimposed on the atmospheric transmission at Mauna Kea.

\section{Photometric properties}

Using a modified version of the ATRAN program (Lord 1992), we
constructed models of the telluric absorption as a function of air
mass to investigate the photometric properties of the new filter set. 
Absorption spectra covering the range 0.8--5.0 $\mu$m were generated
for 10 zenith angles between 0$^{\circ}$ and 70$^{\circ}$ for each of
five values of the precipitable water vapor content (0.5, 1.0, 2.0,
3.0, and 4.0 mm).  The zenith angles correspond to a range of air
masses between 1.0 and 2.9.  The values of the water vapor content
span the range typically found on Mauna Kea (see, e.g., Morrison et
al.  1973; Warner 1977; Wallace \& Livingston 1984).  To examine the
photometric characteristics down to an air mass of 0.0, we also
calculated spectra for five additional air mass values $<$1.0 (0.05,
0.10, 0.25, 0.50, and 0.75) for each of the water vapor values.  These
spectra were generated with ATRAN using a model atmosphere in which
both the gas and water vapor content were given by the standard model
values (appropriate for Mauna Kea at an air mass of 1) scaled by the
air mass value.  The water content was then scaled by an additional
factor given by the ratio of the desired water vapor value to the
standard model value computed by ATRAN for the atmosphere above Mauna
Kea.  All spectra were computed for an altitude of 4158 m (which is
the altitude of the NASA Infrared Telescope Facility) and with a
resolution of approximately 100,000.

The telluric absorption spectra were then multiplied by a similar
resolution model spectrum of Vega obtained from R. Kurucz (2001).  For
each filter, we computed synthetic magnitudes by multiplying the
resulting spectra by the normalized filter transmission curves and
integrating over the wavelength range spanned by the filter. 
Magnitude differences relative to the original model spectrum of Vega
were calculated.  In Fig.  3 we show the magnitude differences versus
air mass for the case of 2.0 mm precipitable water.

For each filter and water vapor value, the run of magnitude
differences as a function of air mass was fitted with a least-squares
routine with two functional forms: (1) a linear fit between air masses
of 1 and 3 and (2) the rational expression given by Young et al. 
(1994; see their eq.  7) for the entire range of air masses.  The
latter expression has the form

\begin{equation}
     \Delta m = (a + b X + c  X^2) / (1 + d X)
\end{equation}

\noindent where $a$, $b$, $c$, and $d$ are fitting constants and $X$
is the air mass.  A detailed discussion of this function is given by
Young (1989).  Both fits are shown in Fig.  3, and the coefficients
are given in Tables 2 and 3.

The linear fit simulates the typical photometric reduction technique 
employed by ground-based observers to obtain the extinction 
coefficient in magnitude per air mass.  The constant term (the 
intercept of the linear fit) yields the systematic error in the 
photometric magnitude incurred by adopting this particular functional 
form of the atmospheric extinction and extrapolating to zero air mass.  
Inspection of the values given in Table 2 indicates that systematic 
errors $>$0.05 mag result from a linear extrapolation for 
the $K'$ and $M'$ filters for all values of the water vapor.  The 
errors are decreased only slightly if the fit is restricted to the 
air mass range 1--2.

As can be seen in Fig.  3, equation (1) provides an excellent fit to 
all the data points.  However, the determination of the four 
coefficients of such a function from actual observations of standard 
stars is impractical for most observing programs, as a large number of 
measurements over a range of air masses are required.  Even if these 
observations were carried out, reliable values can be obtained only if 
measurements in the completely inaccessible air mass range between 0 
and 1 were available.  The values in Table 3 are presented for 
comparison to other photometric filters, such as those discussed by 
Young et al. (1994).

The proposed filters show greatly improved photometric properties over
older filter sets.  Krisciunas et al.  (1987) summarized the average
extinction coefficients at Mauna Kea.  Comparison of our values given
in column 3 of Table 2 with those given in Table I by Krisciunas et
al.  (1987) reveals that at the typical water vapor of 2--3 mm of
precipitable water for Mauna Kea, the reduction in the extinction
coefficient is about a factor of 8 at J. In addition, comparison of
the values given in column 2 of Table 2 with those given in Table III
of Manduca \& Bell (1979) shows that the reduction in the error of
extrapolation to zero air mass (which they denote as $\Delta$) is
about a factor of 9 at J compared to the KPNO winter values.

As was pointed out by Manduca \& Bell (1979), the extinction error 
upon extrapolation to zero air mass does not affect differential 
photometry, in which one compares observed fluxes of target objects to 
those of established standard stars.  The extinction errors in this 
case will be absorbed into the photometric zero points.  It should be 
noted, however, that the extinction error may vary with the 
temperature of the object being observed.  This variation leads to a 
color term in the extinction coefficient.  However, as demonstrated by 
Manduca \& Bell, this term is generally small, amounting to an error 
of $<$0.001 mag for most filters.  Only the $J$ band filter 
analyzed by Manduca \& Bell exhibited a significant color dependence.  
Given that the new $J$ filter is less affected by the atmospheric 
absorption, we expect that the extinction color term for this filter 
will be as small as those for the other filters.

The extinction errors introduced by adopting a linear fit may,
however, affect absolute photometric measurements as, for example, in
the establishment of an exo-atmospheric magnitude system.  The latter
subject has a number of additional complications beyond the scope of
the present paper.  We simply wish to point out that such a system
requires some way to eliminate the extinction error (for example, with
narrowband filters; see Young et al.  1994).

\section{Summary}

A set of 1--5 $\mu$m filters designed for improved photometry and for 
use with adaptive optics is described.  Although the filter profiles are 
optimized for maximum throughput, they avoid most of the detrimental 
effects of atmospheric absorption bands.  A production run of 
these filters has been completed.  It is our hope that widespread 
use of these filters will help produce more uniform photometric 
results, reduce the magnitude of the color transformation among 
observatories, and reduce the uncertainty in reductions to zero or 
unit air mass.

\acknowledgments 

We thank E. Atad, J. Elias, S. Leggett, K. Matthews, and especially T.
Hawarden for discussions and input on the filter specifications.  We
also thank S. Lord for providing a modified version of ATRAN that
allowed the calculation of the atmospheric transmission between 
air masses of 0.0 and 1.0.  Filter profiles in electronic form may be 
found at {http://irtf.ifa.hawaii.edu/Facility/nsfcam/hist
/newfilters.html}. 
ATT acknowledges the support of NASA Contract NASW-5062 and NASA Grant
No.  NCC5-538.  DS acknowledges the support of the Gemini Observatory,
which is operated by the Association of Universities for Research in
Astronomy, Inc., under a cooperative agreement with the NSF on behalf
of the Gemini partnership: the National Science Foundation (United
States), the Particle Physics and Astronomy Research Council (United
Kingdom), the National Research Council (Canada), CONICYT (Chile), the
Australian Research Council (Australia), CNPq (Brazil), and CONICET
(Argentina).


\clearpage





\vspace{10mm} \noindent Fig.  1(a)--(g).  Filter profiles at a
temperature of 65 K. These were measured by the vendor at room
temperature, then shifted according to the measured change with
temperature of a witness sample.

\noindent Fig.  2.    $J, H, K_s, L', M'$ filter profiles 
superimposed on the atmospheric transmission at Mauna Kea  kindly 
provided by G. Milone for 1 mm precipitable water vapor and an 
air mass of 1.0.

\noindent Fig. 3.  Extinction plot for 2.0 mm precipitable water.  
The dashed lines are the linear fits to the magnitudes in the air mass 
range of 1.0--3.0.   The solid lines are the fit to equation (1) in 
the air mass range of 0.0--3.0.

\clearpage


\begin{figure}
\plotone{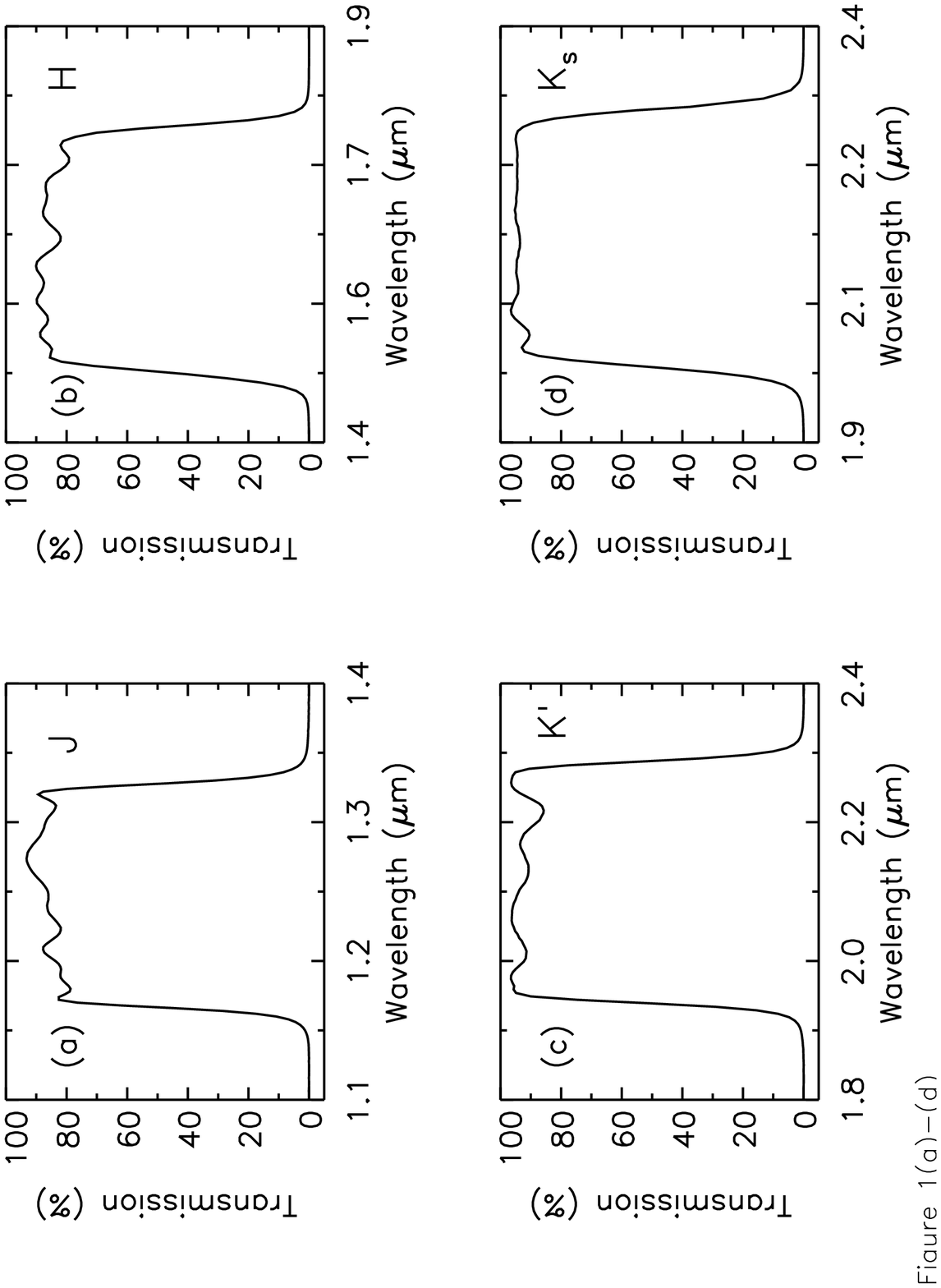}
\end{figure}

\clearpage

\begin{figure}
\plotone{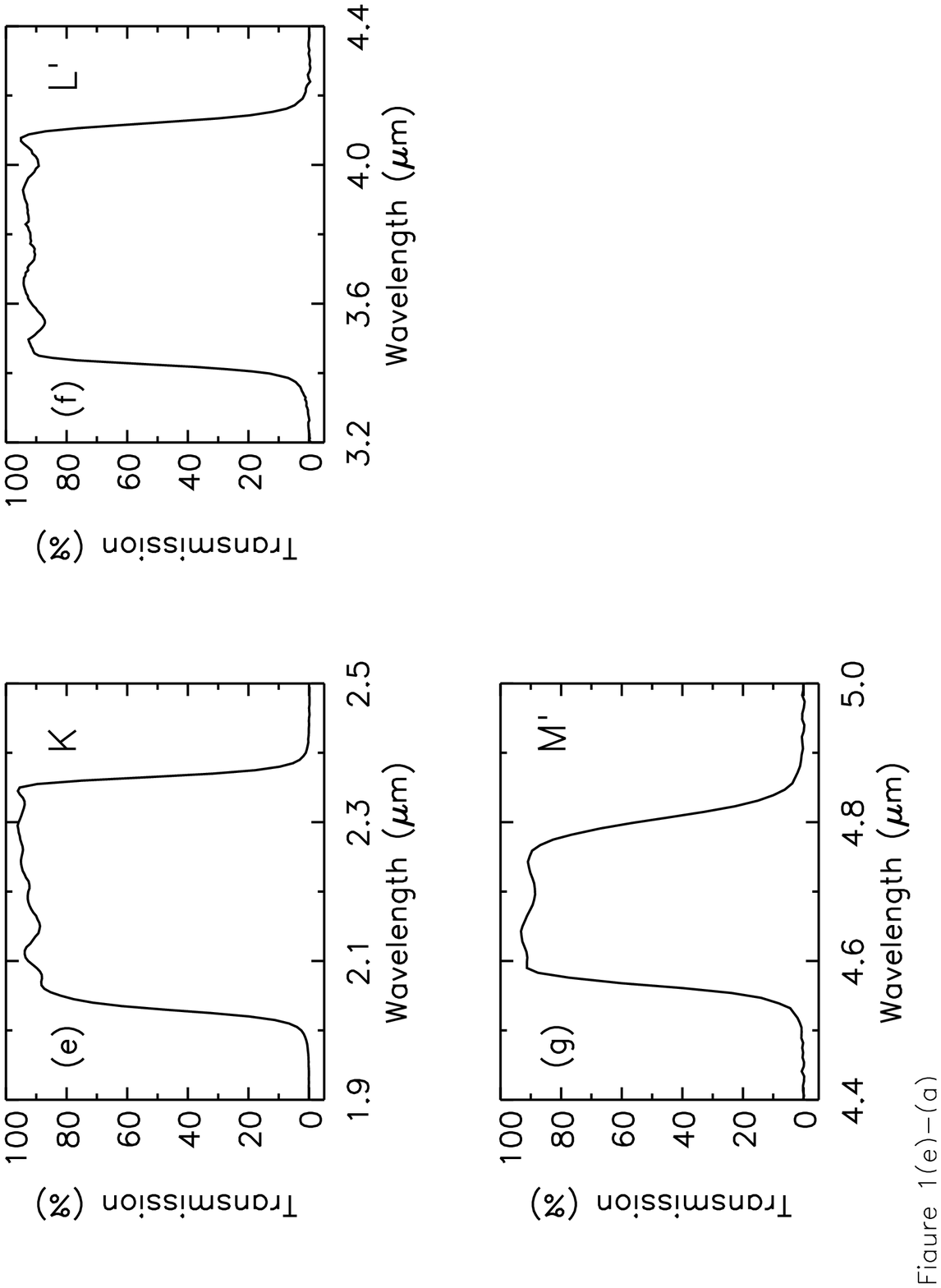}
\end{figure}

\clearpage

\begin{figure}
\plotone{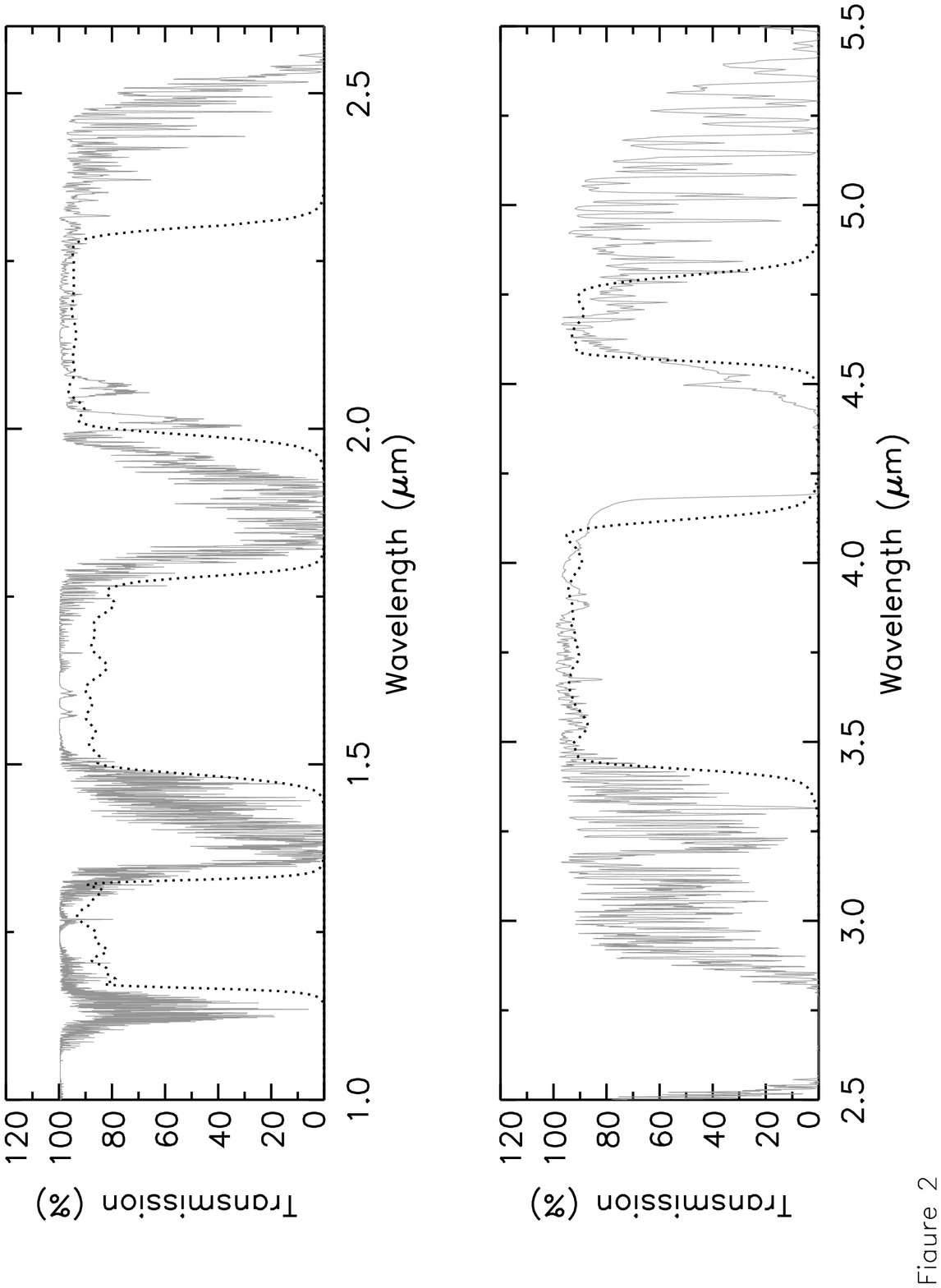}
\end{figure}

\clearpage

\begin{figure}
\plotone{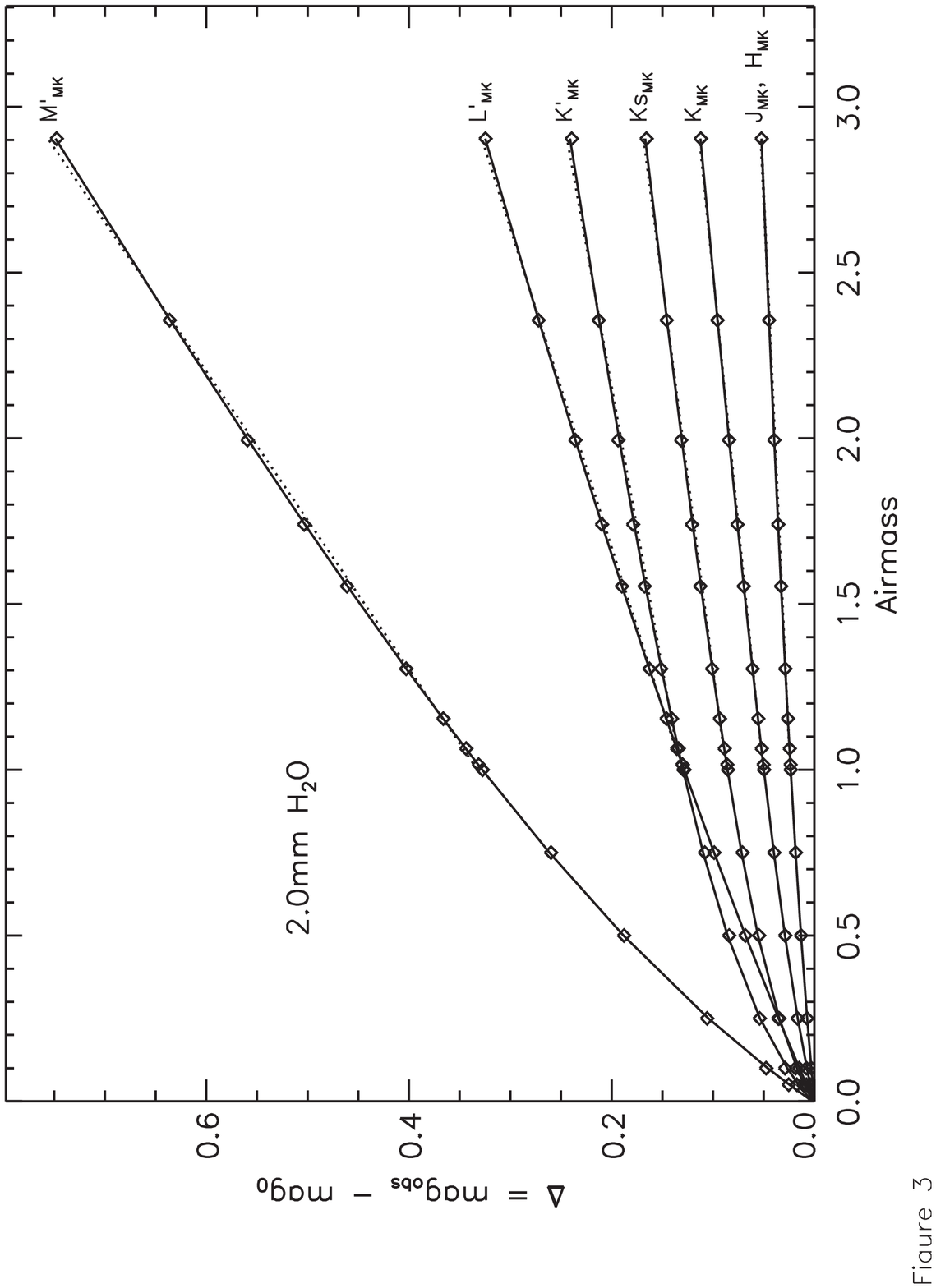}
\end{figure}

\clearpage

\begin{deluxetable}{crrr}
\tabletypesize{\scriptsize}
\tablecaption{Filter Center, Cut-on, and Cut-off Wavelengths \label{tbl-1}}
\tablewidth{0pt}
\tablehead{
\colhead{Name} & \colhead{Center}   & \colhead{Cut-on}   &
\colhead{Cut-off} }
\startdata
$J$       & 1.250 & 1.170 & 1.330 \\
$H$       & 1.635 & 1.490 & 1.780 \\
$K'$     & 2.120 & 1.950 & 2.290 \\
$K_s$ & 2.150 & 1.990 & 2.310 \\
$K$       & 2.200 & 2.030 & 2.370 \\
$L'$    & 3.770 & 3.420 & 4.120 \\
$M'$    & 4.680 & 4.570 & 4.790 \\

\enddata

\tablecomments{The cut-on and cut-off wavelengths correspond to 
where the transmission is 50\% of the peak.}

\end{deluxetable}

\clearpage

\begin{deluxetable}{crr}
\tabletypesize{\scriptsize}
\tablecaption{Linear Fit to Calculated Extinction Curves\label{tbl-1}}
\tablewidth{0pt}
\tablehead{
\colhead{ mm H$_{2}$O/} & \colhead{Constant}   & \colhead{Slope}  \\
\colhead{ filter } & \colhead{ (mag) }   & \colhead{ (mag) }  }
\startdata
0.5 mm &   &  \\
$J$         &  0.0043   $\pm$ 0.0002    &  0.0076   $\pm$ 0.0001 \\
$H$        &  0.0030   $\pm$ 0.0002    & 0.0088   $\pm$ 0.0001 \\
$K$        & 0.0170 $\pm$ 0.0008     & 0.0288   $\pm$ 0.0004 \\
$K'$   & 0.0574  $\pm$ 0.0021    & 0.0491   $\pm$ 0.0012 \\
$K_s$ & 0.0429  $\pm$ 0.0016    & 0.0384   $\pm$ 0.0009 \\
$L'$    & 0.0263   $\pm$ 0.0019    & 0.0938   $\pm$ 0.0011 \\
$M'$  & 0.0956   $\pm$ 0.0042    & 0.2020   $\pm$ 0.0025 \\
1.0 mm &   &  \\
$J$         &  0.0056  $\pm$ 0.0003    &  0.0107   $\pm$ 0.0002  \\
$H$        &  0.0053   $\pm$ 0.0003   &  0.0114   $\pm$ 0.0002 \\
$K$        & 0.0171   $\pm$ 0.0008    & 0.0304   $\pm$ 0.0004  \\
$K'$   & 0.0637   $\pm$ 0.0023    & 0.0535   $\pm$ 0.0014  \\
$K_s$ & 0.0433   $\pm$ 0.0016   &  0.0401   $\pm$ 0.0009  \\
$L'$    & 0.0261   $\pm$ 0.0019    & 0.0975   $\pm$ 0.0011  \\
$M'$  & 0.1009   $\pm$ 0.0045    & 0.2109   $\pm$ 0.0026 \\
2.0 mm &   &  \\
$J$         &  0.0085   $\pm$ 0.0005   &  0.0153   $\pm$ 0.0003  \\
$H$        &  0.0091   $\pm$ 0.0005    & 0.0149   $\pm$ 0.0003 \\
$K$        & 0.0175   $\pm$ 0.0008    & 0.0331   $\pm$ 0.0005 \\
$K'$   & 0.0731   $\pm$ 0.0027  &   0.0589   $\pm$ 0.0015  \\
$K_s$ & 0.0442   $\pm$ 0.0017   &  0.0429   $\pm$ 0.0010  \\
$L'$    & 0.0264   $\pm$ 0.0020   &  0.1039   $\pm$ 0.0012  \\
$M'$  & 0.1099   $\pm$ 0.0049  &   0.2226   $\pm$ 0.0028 \\
3.0 mm &   &  \\
$J$         &  0.0115   $\pm$ 0.0006   &  0.0188   $\pm$ 0.0004 \\
$H$        &  0.0124   $\pm$ 0.0006  &   0.0173   $\pm$ 0.0003 \\
$K$        & 0.0180   $\pm$ 0.0008   &  0.0354   $\pm$ 0.0005 \\
$K'$   & 0.0803   $\pm$ 0.0029   &  0.0625   $\pm$ 0.0017  \\
$K_s$ & 0.0451   $\pm$ 0.0017   &  0.0452   $\pm$ 0.0010  \\
$L'$    & 0.0274   $\pm$ 0.0021   &  0.1094   $\pm$ 0.0012  \\
$M'$  & 0.1170   $\pm$ 0.0051   &  0.2313   $\pm$ 0.0030 \\
4.0 mm &   &  \\
$J$         &  0.0144  $\pm$ 0.0008  &   0.0215  $\pm$ 0.0004\\
$H$        &  0.0151   $\pm$ 0.0007   &  0.0191   $\pm$ 0.0004\\
$K$        & 0.0185  $\pm$ 0.0009   &  0.0375   $\pm$ 0.0005 \\
$K'$   & 0.0858   $\pm$ 0.0030   &  0.0653   $\pm$ 0.0018  \\
$K_s$ & 0.0461   $\pm$ 0.0017   &  0.0473   $\pm$ 0.0010  \\
$L'$    & 0.0288   $\pm$ 0.0022   &  0.1140   $\pm$ 0.0013  \\
$M'$  & 0.1225   $\pm$ 0.0053  &   0.2380   $\pm$ 0.0031 \\

\enddata

\tablecomments{Least-squares linear fitting to the equation 
(const. + slope*X), where X is the air mass in the range 1.0--3.0.}

\end{deluxetable}


\begin{deluxetable}{crrrr}
\tabletypesize{\scriptsize}
\tablecaption{Fit to Calculated Extinction Curves using Eq. (1)\label{tbl-1}}
\tablewidth{0pt}
\tablehead{
\colhead{ Filter } & \colhead{a}   & \colhead{b} & \colhead{c}   & \colhead{d}  }
\startdata
0.5 mm &   &  \\
$J$         &  0.0000  & 0.0166  & 0.0047  & 0.8166 \\
$H$        &  0.0000  & 0.0146  & 0.0043  & 0.6085 \\
$K$        & 0.0005  & 0.0710  & 0.0301   & 1.2459 \\
$K'$   & 0.0018  & 0.2376  & 0.0819   & 2.0629 \\
$K_s$ & 0.0016  & 0.1782   & 0.0645  & 2.0490 \\
$L'$    & 0.0006  & 0.1347  & 0.0223  & 0.3311 \\
$M'$  & 0.0013  & 0.4301  & 0.1953  & 1.1315\\
1.0 mm &   &  \\
$J$         &  0.0000  & 0.0213  & 0.0050  & 0.6464  \\
$H$        &  0.0001  & 0.0210  & 0.0047  & 0.5642 \\
$K$        & 0.0005  & 0.0718  & 0.0306   &1.1987  \\
$K'$   & 0.0022  & 0.2507  & 0.0796  & 1.8836  \\
$K_s$ & 0.0016  & 0.1793  & 0.0663  & 2.0116 \\
$L'$    & 0.0006  & 0.1378  & 0.0228  & 0.3237  \\
$M'$  & 0.0017  & 0.4402  & 0.1829   &1.0317\\
2.0 mm &   &  \\
$J$         &  0.0001  & 0.0310  & 0.0061  & 0.5891 \\
$H$        &  0.0002  & 0.0331  & 0.0082  & 0.7624 \\
$K$        & 0.0006  & 0.0738  & 0.0306   &1.1041\\
$K'$   & 0.0026  & 0.2828  & 0.0851  & 1.8568  \\
$K_s$ & 0.0017  & 0.1813  & 0.0680  & 1.9290  \\
$L'$    & 0.0006  & 0.1441  & 0.0224  & 0.2998  \\
$M'$  & 0.0020 &  0.4691  & 0.1866   &1.0066 \\
3.0 mm &   &  \\
$J$         &  0.0001  & 0.0407  & 0.0082  & 0.6464 \\
$H$        &  0.0003  & 0.0449  & 0.0130  & 0.9979 \\
$K$        & 0.0006  & 0.0770  & 0.0322   &1.0802\\
$K'$   & 0.0028  & 0.3163  & 0.0954   &1.9525 \\
$K_s$ & 0.0017  & 0.1851  & 0.0706  & 1.8917  \\
$L'$    & 0.0006  & 0.1509  & 0.0227  & 0.2908 \\
$M'$  & 0.0021  & 0.5013  & 0.2067  & 1.0670 \\
4.0 mm &   &  \\
$J$         &  0.0001  & 0.0506  & 0.0113  & 0.7586\\
$H$        &  0.0003  & 0.0577  & 0.0198   &1.3120\\
$K$        & 0.0005  & 0.0807  & 0.0345   &1.0899\\
$K'$   & 0.0028  & 0.3523  & 0.1100  & 2.1260 \\
$K_s$ & 0.0017  & 0.1896  & 0.0735   & 1.8790 \\
$L'$    & 0.0006  & 0.1580  & 0.0244  & 0.2986 \\
$M'$  & 0.0019  & 0.5381  & 0.2425  & 1.1978 \\

\enddata

\tablecomments{ Least-squares fitting to equation (1) in the 
air mass range of 0.0--3.0.}

\end{deluxetable}

\end{document}